\documentclass[journal]{IEEEtran}
\usepackage{amsmath,amssymb,amsfonts}
\usepackage{physics}
\usepackage{algorithmic}
\usepackage{graphicx}
\usepackage{textcomp}
\usepackage[style=ieee,isbn=false,url=false,eprint=false,backend=bibtex]{biblatex}
\usepackage{soul}
\begin{document}

\title{Trusted Node QKD at an Electrical Utility}
\author{Philip~G.~Evans~[1]*,~\IEEEmembership{Member,~IEEE,}
        Muneer~Alshowkan~[1],~\IEEEmembership{Member,~IEEE,}
        Duncan~Earl~[2],
        Daniel~Mulkey~[2],
        Raymond~Newell~[3],
        Glen~Peterson~[3],
        Claira~Safi~[3],
        Justin~Tripp~[3],~\IEEEmembership{Member,~IEEE}
        and~Nicholas~A.~Peters~[1],~\IEEEmembership{Senior~Member,~IEEE}

\thanks{* Corresponding author: evanspg@ornl.gov}
\thanks{[1] Quantum Information Science Group, Computational Sciences \& Engineering Division, Oak Ridge National Laboratory, Oak Ridge, TN 37831}
\thanks{[2] Qubitekk Inc., Vista, CA 92081}
\thanks{[3] Los Alamos National Laboratory, Los Alamos, NM 87545}
\thanks{This manuscript has been co-authored by UT-Battelle, LLC, under contract DE-AC05-00OR22725 with the US Department of Energy (DOE). The US government retains and the publisher, by accepting the article for publication, acknowledges that the US government retains a nonexclusive, paid-up, irrevocable, worldwide license to publish or reproduce the published form of this manuscript, or allow others to do so, for US government purposes. DOE will provide public access to these results of federally sponsored research in accordance with the DOE Public Access Plan (http://energy.gov/downloads/doe-public-access-plan).}}

\maketitle
\begin{abstract}
Challenges facing the deployment of quantum key distribution (QKD) systems in critical infrastructure protection applications include the optical loss-key rate tradeoff, addition of network clients, and interoperability of vendor-specific QKD hardware. Here, we address these challenges and present results from a recent field demonstration of three QKD systems on a real-world electric utility optical fiber network.
\end{abstract}


\section{Introduction}
\label{sec:introduction}
Electric power is a fundamental utility in modern society. For the most part, today's power systems rely on dated technology developed for one-way power flows from large power plants to passive customers. In the United States and many other countries, modernization of the electric power grid is central to efforts aiming to increase reliability and energy efficiency, transitioning to renewable sources of energy, reducing greenhouse gas emissions, and building a sustainable economy that ensures prosperity for future generations. The \textit{Smart Grid} is not new grid infrastructure - it is the integration of existing power infrastructure with an information infrastructure, thereby combining the maturity of the electric grid with the efficiency, connectivity, and cost gains brought about by Information Technology (IT) \cite{Budka2010, Gungor2013}. However, enormous challenges remain in developing and integrating Smart Grid elements to realize the next generation of Supervisory Control and Data Acquisition Systems (SCADA) for Operational Technology (OT) applications \cite{Sridhar2012, Tomsovic2005}.

Data collection, analysis and decisions are central to management of the smart grid. Data, collected by a variety of sensors, comprise the usual power systems metrics of voltage, current, frequency, and phase of the grid at a specific geographic point; but also include meteorological data pertinent for renewable energy generation such as wind speed and direction, solar flux and cloud cover, and instantaneous charge available on battery banks \cite{Zhou2016}. Following analysis of this data, decisions are made by humans-in-the-loop - but also, increasingly, by machine-based decision makers - leading to commands issued which implement controls on the smart grid itself. The communications flow in the smart grid is assymetric: the majority of information consists of sensor data flowing \textit{into} the grid operations center from networked sensors. In contrast, only a relatively small amount of information consists of command and control data originating from the operations center \textit{outward} to the power equipment \cite{Tomsovic2005, Ma2013}. All communications on the smart grid have the same requirements: it must be both trustworthy and timely \cite{Metke2010, Wang2011, Gungor2013}. The operations center must intrinsically trust the accuracy and validity of the data it receives from grid sensors, whereas the power equipment must trust commands received from the operators. Similarly, it is vitally important sensor data and the resultant grid configuration commands reach the intended recipient with minimal communications latency. The smart grid requires a foundational communications security infrastructure which can provide for authentication and encryption of messages, yet without the severe computational resource (and thus latency) penalties accompanying public-key cryptographic schemes on limited computer hardware.

With these requirements and constraints in mind, quantum communications - specifically quantum key distribution - provides a unique solution for security of the smart grid, and promises to provide the foundation for SCADA communications in a future-proof manner with security regardless of technological developments. In this paper, we describe the first demonstration of a multiple trusted-node QKD network operating at an electrical utility. This demonstration is the culmination of several years of effort to prove the feasibility of QKD-based security for energy delivery systems. This paper is organized as follows: in Section \ref{sect:QKD} we provide a brief overview of quantum key distribution and why the electric grid provides a compelling use-case for QKD deployments; in Section \ref{sect:demo} we describe the QKD systems, network configuration, and software functionality employed in the demonstration; Section \ref{sect:results} presents results from the trusted-node QKD network demonstration over several days of data collection and analysis; finally we conclude and summarize in Section \ref{sect:conclusion}.

\section{Quantum Key Distribution}
\label{sect:QKD}
Quantum key distribution (QKD) \cite{Bennett2014} is one of the most mature quantum technologies. Briefly, QKD allows two spatially separated parties (typically referred to as \textit{Alice} and \textit{Bob}) to generate and share private strings of random numbers. This \textit{key material} is known only to Alice and Bob, even assuming the communication channels are controlled by an arbitrarily powerful adversary \textit{Eve}, who is bounded only by the laws of physics. For example, any action by Eve in an attempt to clone, tamper, or modify the quantum communications between Alice and Bob are detectable, with Eve gaining no information advantage \cite{Lo1999, Brassard2000, Shor2000}.

A variety of techniques have been proposed and demonstrated for QKD \cite{Gisin2002, Xu2020, Huang2018}. The most mature approaches utilize discrete-variable (DV) quantum encodings on heavily attenuated laser pulses \cite{Buttler2000, Stucki2002, Ma2005, Korzh2015}, single-photons \cite{Alleaume2004, Intallura2007, Takemoto2015} and entangled-photons \cite{Lee2014, Marcikic2006, Ma2007}. An attractive alternative using continuous-variable (CV) encodings, such as amplitude and phase, can be realized through Gaussian modulation of coherent states \cite{Ralph2000, Cerf2007, Jouguet2013}, as well as passively with thermal states \cite{Weedbrook2012b, Qi2020}. While optical fiber remains the physical transmission medium of choice, free-space QKD \cite{Buttler2000, Hughes2002, Erven2008, Peloso2009} as well as satellite-based QKD \cite{Vallone2015, Liao2017a, Liao2017b, Chen2021} has also been demonstrated. QKD has successfully transitioned from lab-grade research experiments to systems being sold and installed by several global companies \cite{IDQ, Qubitekk}. Research into advanced QKD protocols, as well as more general quantum networking protocols, remains an ongoing and active area of research.

While QKD has many advantages, there are some notable disadvantages. First, QKD rates are fundamentally limited over pure loss channels \cite{Takeoka2014, Pirandola2017}, with the final \textit{secret key rate (SKR)} decreasing with the loss between parties. Second, QKD is limited to pairwise key distribution between Alice and Bob only. Adding users to a QKD link is not straight-forward, and requires different protocols such as quantum secret sharing (QSS) \cite{Tittel2001, Grice2015, Grice2019, Williams2019} and quantum digital signatures \cite{Yin2016, An2018}. Third, interoperability between QKD systems operating with dissimilar implementations can only be accomplished using a classical layer. Only recently have some limited QKD standards been defined, notably by the ETSI Quantum-Safe Cryptography Working Group \cite{ETSI}, in addition to QKD network and QKD systems activities within ITU-T SG13 \cite{ITU-SG13} and SG17 \cite{ITU-SG17} respectively, yet much work remains before standards provide an industry-accepted framework for wide implementation.

\subsection{QKD on the Smart Grid}
Despite these disadvantages, the Smart Grid is a natural platform for QKD and related quantum communications technologies. As highlighted earlier, the machine-to-machine (M2M) communications at the heart of Smart Grid SCADA network is not bandwidth intensive, yet these commands must be authentic. Communications on the grid are not typically between peers - substations rarely need to communicate on a peer-level with other substations and power distribution sites \cite{Ma2013}. The communications are hierarchical, with a central authority issuing commands to operate the grid. Consequently, the authenticity of the command messages must be established. The most stringent method to accomplish this is by encrypting the messages with keys which only the authority and intended recipients have.

Communication latency is also important for grid operations. In a fault event, or any situation where quick action must be taken to avert an outage, commands must be sent, received, and executed with minimum delay \cite{Kansal2012}. The use of common IT data encryption methods, for example public-key cryptography based on Rivest-Shamir-Adleman (RSA) or equivalent algorithms, or using certificates, requires greater computational resources than with private-key cryptography. Thus, the most computationally efficient (and therefore lowest latency) method remains the one-time pad (OTP) method, where a message is combined with the exclusive OR operation (XOR) and the key. OTP exhibits information theoretical security (ITS), that is to say it is secure regardless of an adversary's computational power, with the following caveats: (1) the keys must be truly random, be kept secret, and are used once only, and (2) the message length is less than or equal to the length of the key \cite{Shannon1948}.

Electrical grids operated by a single utility or cooperative have a relative small geographic footprint. Their service area is often metropolitan, with most substations being within 25 km of another \cite{HIFLD}. Due to the cost of radio spectrum, utilities are increasingly deploying optical fiber as part of their smart grid investment, for example, between operations centers, substations, large industrial users, and distributed energy resources. At the same time, infrastructure investments are expected to have a long service life, creating the need for long-term cybersecurity. Further, conductor loss creates an incentive to minimize the length of distribution systems, which is a similar constraint for QKD networks. 

The above requirements for long-term security, limited bandwidth, low latency, limited computational resources, a hierarchical network topology, relatively short metropolitan-scale links, and the capability of deploying optical fiber along existing electrical conductor right-of-ways, constitute a use-case where QKD is a well-matched solution.

Any QKD-based grid security solution must be fully tested and certified for use in an electrical utility environment. The electrical and environmental conditions present in typical substations are remarkably different than those in a laboratory \cite{Hughes2013}. Swings in temperature and humidity, the presence of electromagnetic interference due to high-voltage and high-current switching equipment, differences in power sources and stability (e.g., 48 Volts Direct Current (VDC) supply being inverted for 120 Volts Alternating Current (VAC)), etc., have the potential to disrupt sensitive electronics, optics and electro-optics that comprise QKD hardware systems. In addition, the optical environment must also be carefully considered: optical fiber loss is commonly assumed to be approximately 0.2 dB/km for standard single mode fiber at a wavelength of 1550 nm; yet deployed fiber, complete with various connections can result in an \textit{effective} optical loss figure much larger than the bare fiber alone. Fiber installation methods, such as direct burial, via conduit, aerial, or some combination can also impact achievable secret key distribution rates. 

\subsection{Trusted Relay QKD}
In the QKD context, a \textit{trusted relay} is a type of simple quantum network device where quantum signals are terminated (via measurement) at relay locations. All switching and routing tasks are handled classically on the digitized measurement results. Trusted relay QKD, also known as trusted node QKD, has been realized in several demonstrations, starting with the DARPA Quantum Network \cite{Elliott2002}, along with the SECOQC program \cite{Peev2009}, the Swissquantum network \cite{Stucki2011}, the Tokyo quantum network \cite{Sasaki2011}, several quantum networks recently demonstrated in China \cite{Chen2010, Wang2014, Mao2018}, and the Cambridge quantum network \cite{Dynes2019}. In 2019, we completed the first steps to establish the feasibility of trusted relay QKD on a utility network using a single trusted location with a fiber loop-back \cite{Evans2019}. Building upon our earlier conference paper, here we report the first utility-based quantum network. Specifically, our network utilizes three different types of QKD systems located at four trusted nodes at different locations in EPB's Utility infrastructure in Chattanooga Tennessee.

\begin{figure}[!t]
\centering
\includegraphics[width=3.3in]{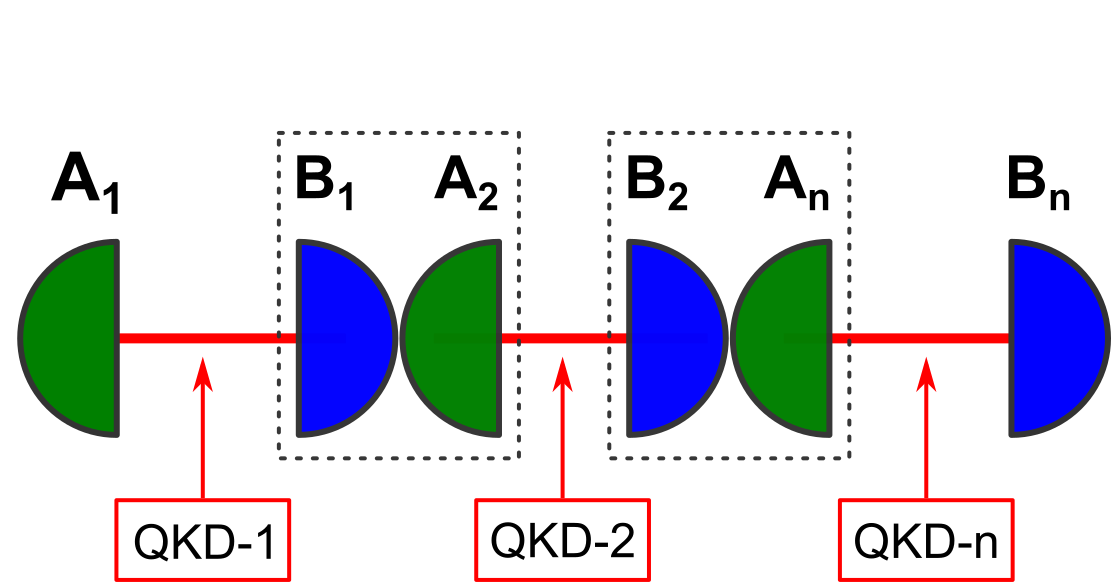}
\caption{Illustration of a linear trusted relay QKD network. Alice-n ($A_{(n)}$) and Bob-n ($B_{(n)}$) are connected by QKD link $n$. The intermediate nodes $B_{(n-1)}A_{(n)}$ are physically co-located (for example, in a substation) and comprise the trusted relay. The required classical communications channel is not shown.}
\label{fig1}
\end{figure}

Trusted relay QKD allows the distance limitation inherent in a single QKD link to be overcome. It also allows for a mechanism to add users, and in essence, the creation of a QKD network for key exchange. Trusted node measurement of single photons converts quantum information to classical information and as a result enables classical-layer interoperability between disparate underlying QKD hardware. Recall QKD is a pairwise communications scheme between \textit{Alice-N ($A_{(n)}$)} and \textit{Bob-N ($B_{(n)}$)}. In a linear network configuration, the start and end nodes are $A_{(1)}$ and $B_{(n)}$ respectively. The intermediate nodes $B_{(n-1)}A_{(n)}$ are physically co-located yet perform QKD on distinct quantum links - these are the trusted relay nodes. $QK_{(n)}^{i}$ is the $i^{th}$ QKD-generated secret key on link $n$. This concept is illustrated in Fig. \ref{fig1}.

Secure key distribution on a trusted-relay QKD network is accomplished by several methods, the most common being the \textit{hop-by-hop} scheme \cite{Peev2009, Sasaki2011}. Here, messages $M^{h}$ originating at $A_{(1)}$ are encrypted using XOR OTP with QKD keys $i$ for link $n = 1$, producing the encrypted message $EM_{(1)}^{h,i}$ i.e., $EM_{(1)}^{h,i} = M^{h} \oplus QK_{(1)}^{i}$. $EM_{(1)}^{h,i}$ is subsequently transmitted over the classical channel to the node $B_{(1)}$ where $M^{h}$ is recovered by performing the reverse operation, i.e., $M^{h} = EM_{(1)}^{h,i} \oplus QK_{(1)}^{i}$, with $QK_{(1)}^{i}$ having been established by QKD. At the trusted relay, $M^{h}$ is now re-encrypted with the QKD key established for $n = 2$, i.e., $EM_{(2)}^{h,j} = M_{h} \oplus QK_{(2)}^{j}$ for passing to the next node. This process of node hopping progresses until the end node $B_{(n)}$ receives the original message. Note the distinction between QKD key indices $i$ (for link $n = 1$) and $j$ (for link $n = 2$) and so forth for all $n$, such that $i \neq j$, to account for any discrepancies in key indices related to physical link conditions and QKD system key generation rates.

It is now apparent why the intermediate nodes need to be trusted - the messages are clear text during the decryption/re-encryption process in the trusted part. Electrical substations, with existing physical security and surveillance including fences, cameras, walls, the watchful eye of the operator, etc., offer physical protection of critical components including QKD relay nodes.

In our demonstration, we wish to securely generate and distribute a pool of \textit{network keys (NKs)}, similar to the PKI key in Ref.\cite{Stucki2011}, amongst substations comprising the QKD network. These network keys enable secure command and control communications between the network operations center and electrical substations. The network keys $NK^{h}$ comprise our messages in hop-by-hop method. $NK^{h}$ are produced at $A_{(1)}$ from a stand-alone quantum random number generator (QRNG) distinct from that used in the QKD systems.

\section{Electrical Utility Trusted Relay Demonstration}
\label{sect:demo}
In this section we describe the configuration of the trusted relay QKD demonstration in collaboration with the Electric Power Board (EPB) in Chattanooga, TN. We describe the QKD systems used, the network configuration, and the software written to perform network key distribution and network monitoring.

The trusted node network consists of the following trusted nodes: a network manager (NM), trusted nodes (TNs), and an end node (EN). The description of each is as follows:

\begin{itemize}
  \item[] \textbf{The network manager (NM)} is responsible for network management, network key generation and key distribution to all nodes on the network. The NM collects network statistics and performs basic key management functions. The NM houses the QRNG - the source of network keys to be distributed.
  \item[] \textbf{The trusted nodes (TNs)} distribute, by the hop-by-hop method, the network keys to the edge node(s). The TNs are responsible for reporting the quantum key statistics to the NM.
  \item[] \textbf{The edge nodes (ENs)} are nodes existing on the network periphery. On a linear network the EN is the final node having only one link. ENs also report the quantum key statistics to the NM.
\end{itemize}

\subsection{QKD Systems}
Three QKD systems were used in the March 2020 demonstration, which we shall refer to as QKD System 1, 2, and 3. QKD System 1 is a commercial system employing the BBM92 protocol \cite{Bennett1992}. QKD System 2 is an advanced research system employing polarization BB84 \cite{LANL1,LANL2}. Finally, QKD System 3 is a commercial system employing phase SARG04 \cite{Scarani2004}. Each system is unique with respect to which protocol and physical encoding is used to perform QKD, and as a result, are not plug and play compatible across different Alice-Bob pairs. Similarly, each QKD system has relative strengths and weaknesses with respect to the possible deployment environments.

\subsection{Network Configuration}
EPB provided two dedicated fiber pairs between each trusted node (six dark fibers total) on their optical fiber network for research and demonstration purposes. Aside from short spans in underground conduit, all fibers used in this demonstration are deployed aerially, i.e., hanging between poles, and thus experience temperature and wind effects accordingly. While these fibers are accessible in electrical substations and other facilities, and mostly run in the same cables, they are distinct from the operational network fibers such that there could be no possibility to affect grid operations. One fiber pair was used to establish a local TCP/IP network for classical communications between QKD systems and to allow remote access; the second fiber pair was utilized for the quantum channel(s) between the respective QKD systems. The research network spans approximately 22 km ($\approx$88 km of dedicated total fiber), beginning at the operations center (OC) and providing connectivity to three major electrical substations A, B, and C, respectively. The OC is a typical data center-type environment with controlled temperature, humidity and clean 120 V AC power. The substations do not have as stringent temperature or humidity controls. In addition, all substation equipment operates from 48 V DC power such that in the event of a power outage, all mission-critical hardware runs from battery backup. In the demonstration described herein, all QKD systems and accompanying hardware ran from the 120 V AC `wall' power; future improvements could integrate suitable DC - AC power inverters, or dedicated 48 V power options, to allow QKD network operations during power outages.

\begin{figure}[!t]
\centering
\includegraphics[width=3.3in]{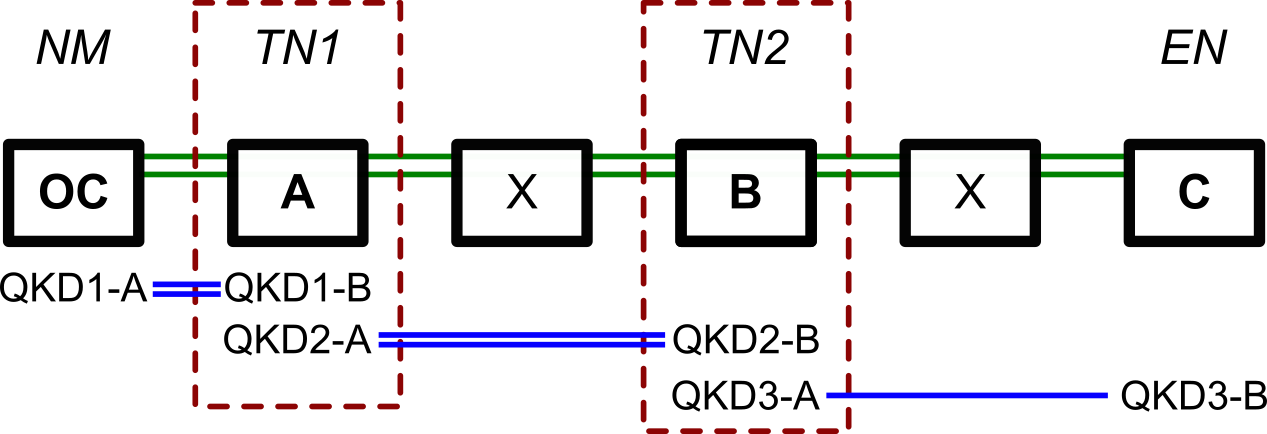}
\caption{Illustration of the EPB Trusted Relay QKD Network topology. The network manager (NM) is located at the operations center (OC); the edge node (EN) located at substation C; trusted nodes (TNs) are located at substations A and B, and are represented with a dashed red outline. Locations X mark substations where access to the fiber is possible, but not used during this demonstration. QKD-x refers to (A)lice and (B)ob hardware, respectively.}
\label{fig2}
\end{figure}

Table \ref{table1} details the network configuration, showing distance and measured optical loss, for the respective QKD system links, between all sites. Fig. \ref{fig2} illustrates the quantum network configuration.

\begin{table}
\caption{Details of the EPB Quantum Network showing optical link length and measured optical loss, as well as the installed QKD system protocol.}
\label{table1}
\begin{center}
\begin{tabular}{|l l l l|}
\hline
Link & Length & Loss & QKD Protocol \\
\hline
\hline
1: OC - A & 3.4 km & 1.3 dB & Polarization BBM92 \\
2: A - B & 10.2 km & 3.1 dB & Polarization BB84 \\
3: B - C & 8.3 km & 2.9 dB & Phase SARG04 \\
\hline
Total & 21.9 km & 7.3 dB & \\
\hline
\end{tabular}
\end{center}
\end{table}

\subsection{Trusted Relay Software}
We have developed software to handle network node operations. Each node in the trusted network is responsible for performing basic functions, such as network communication, providing key statistics, key management, and key transfer, in addition to specialized tasks related to NM, TN, or EN roles. The role of each node and its connection with the other nodes is predefined at launch. For instance, each node knows the NM, the sender and/or the receiver on each channel (for simplicity, we designate the left node as the sender and the right as the receiver) as depicted in Fig. \ref{fig1}. After the initial configuration, the software works autonomously to distribute the key material as it becomes available. We shall briefly describe the common node functionalities.

\subsubsection{Network Communication and Statistics}
Upon software startup, each node creates a connection with the NM and reports the total number of available network and quantum keys, number of used network and quantum keys, together with secret key generation rate (SKR) and quantum bit error rate (QBER) of the link (if available from the QKD systems). The network statistics are stored in a time series database \textit{InfluxDB}) for real-time monitoring and analysis. Network nodes also create a connection with their nearest neighbors, i.e., the left (sender) and right (receiver) nodes in anticipation of network key transfer. Periodically, each node reports a timestamped record of the available number of quantum keys $i, j, k, ...$ and network keys $h$ as QKD runs in the background.

\subsubsection{Key Management}
When a QKD system is running, new key material is being generated at a rate determined by the link conditions and QKD system specifics. The key material is typically written to a file located on both Alice and Bob portions of each node. The node key management function is triggered on events related to key file modification, e.g., when fresh key material is appended, or when stale key material is replaced with fresh key. The event triggers the key updating process to read the key file and process the new key. Careful event handling is needed as each QKD system has a unique way of adding the new key material (appending, re-constructing, or creating a new key file). The key update process handles the key material from different file formats. The key material is retrieved as a string of binary random bit values. Each corresponding node further processes the key material by decoding, chunking, and hashing to create a link-specific key table. Each record in the key table consists of a unique key identifier, a 256-bit key, a SHA-256 hash of the key, and a Boolean status flag indicating use. As the keys are used, they may be removed from the key table, or the status flag set accordingly.

\subsubsection{Key Transfer}
The NM uses the available number of quantum keys per node, i.e., $i, j, k$ for $n = 1, 2, 3$ respectively, to determine how many $NK^{h}$ to transfer. There exist an ample supply of network keys available to transfer from NM, from which, the supply $h$ is drawn. Due to the requirements of strict OTP and distributing network keys to all nodes, $h$ = min\{$i,j,k$\}. We further restrict $h$ by specifying a lower limit on the number of QKD keys available for hop-by-hop key distribution across all nodes. This \textit{reserve pool} of QKD keys can be used, for example, in the event of an emergency where significant disruption of the trusted node network occurs. Thus, $h$ = min\{$i,j,k$\} - $R$ where $R$ is the number of QKD keys kept in reserve network-wide. Finally, we specify a threshold key quantity, $T$, such that $T \geq \{i,j,k\}$ in order to initiate the hop-by-hop network key distribution tasks. In this work, we set $R$ = 20, and $T$ = 60.

The key transfer process starts with the NM accessing the key tables, retrieving $NK^{k}$ and $QK_{(1)}^{k}$, while updating the status flag to indicate key usage. The retrieved keys are used to perform the hop-by-hop scheme to the next node. Each TN retrieves and updates the corresponding $QK^{(n)}_{k}$ for the decryption and encryption operations of the $NK^{k}$ until the key transfer along the network is complete. Each node updates the local $QK^{n}$ key table and stores $NK^{k}$ for future use. The EN notifies the NM upon successful key transfer, ending the key transfer cycle.

\section{Experimental Results}
\label{sect:results}
The following data, as a function of time for each link as well as the overall network, were collected during the field test:
\begin{itemize}
  \item the number of available 256-bit quantum-distributed keys ($QK_{(n)}$),
  \item the number of 256-bit network keys distributed ($NK$),
  \item the secret key generation rate, and
  \item the quantum bit error rate.
\end{itemize}

\subsection{Overall Network Performance}
\begin{figure}[!t]
\centering
\includegraphics[width=3.3in]{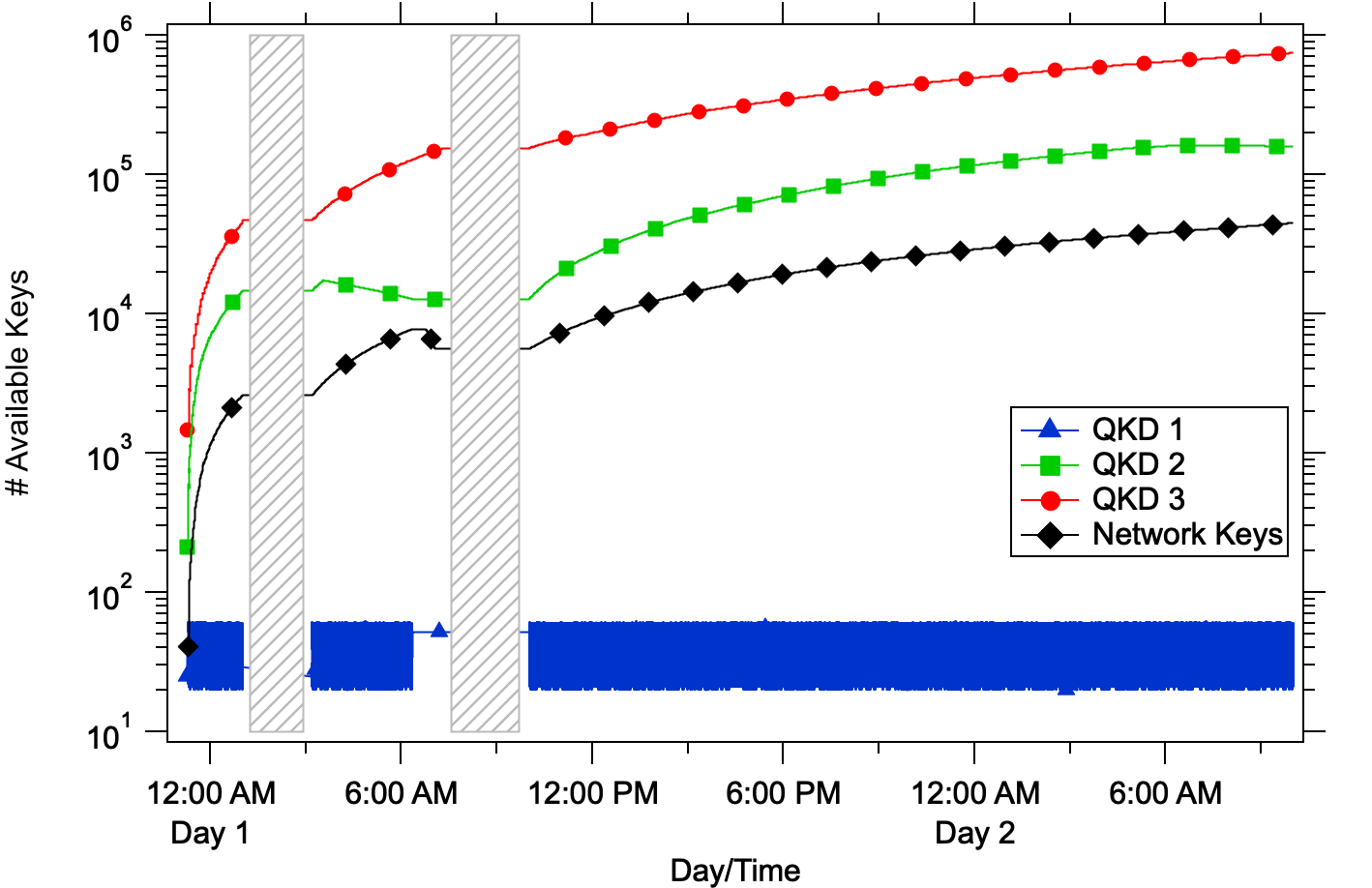}
\caption{Number of available keys as a function of time during the utility field test. The red, green, and blue curves represent the number of available quantum keys $QK_{(n)}$ generated by QKD System $n$, respectively. The black curve represents the number of available network keys ($NK$) following the key transfer cycle.}
\label{fig3}
\end{figure}

Fig. \ref{fig3} shows the number of keys available as a function of time during the field test. The number of available quantum keys for each system link, as well as the number of available network keys distributed over the entire network are shown.

\begin{figure}[!t]
\centering
\includegraphics[width=3.3in]{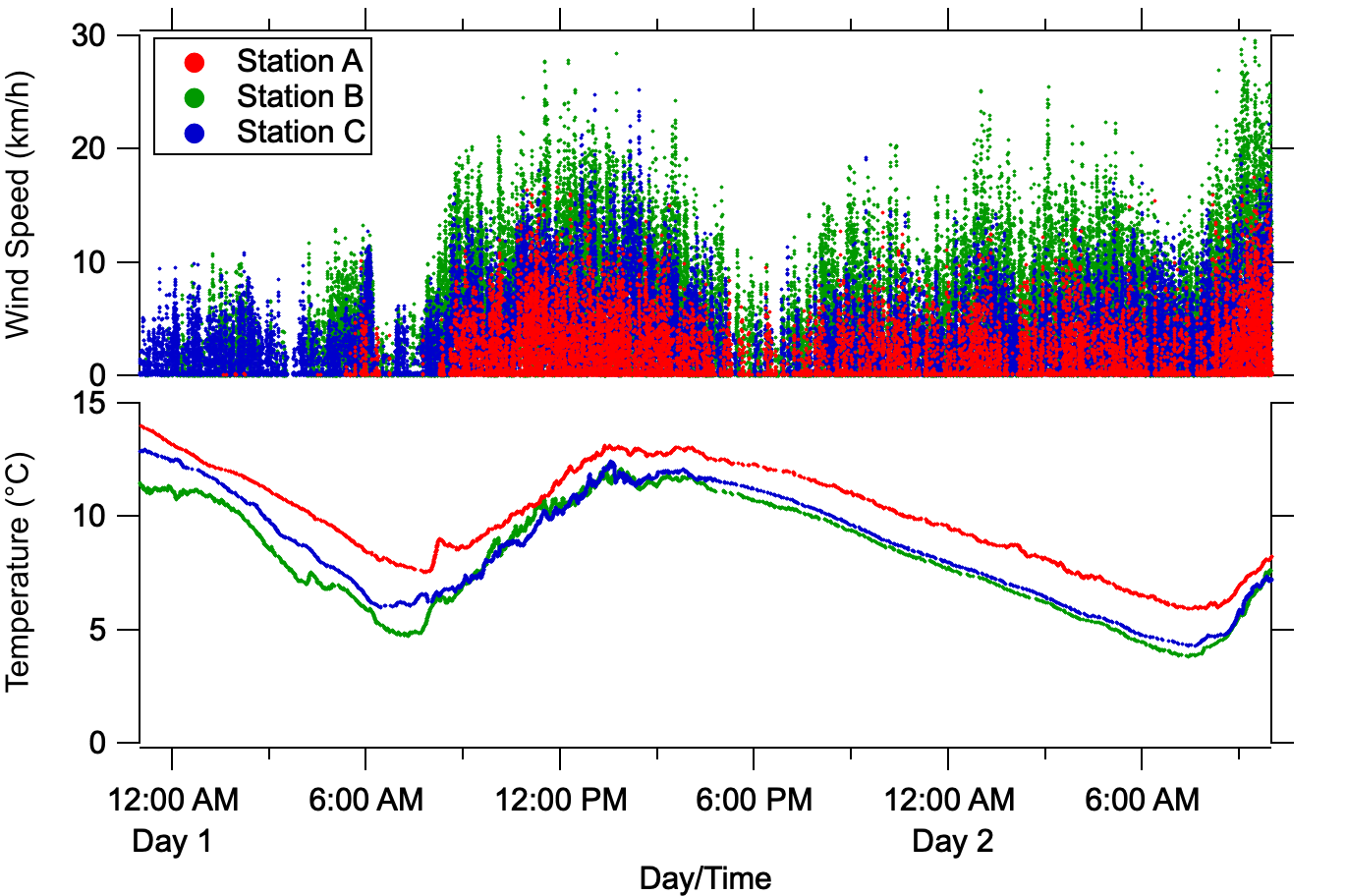}
\caption{Temperature (bottom panel) and wind speed (top panel) recorded at substations A, B, and C during the trusted node demonstration.}
\label{fig4}
\end{figure}

Data is represented as a log-linear plot in Fig. \ref{fig3} to aid the visualization of differing key generation rates over the constituent QKD system links. Each of the key generation rates are approximately constant over time, which shall be covered in detail below. Two distinct gaps appear in the data collection: for approximately 2 1/2 hours beginning 0100, and then for approximately 3 hours beginning 0700, on Day 1. The cause for both occurrences was loss of classical network connectivity between the nodes and NM due to poorly timed automatic software updates. Other observable artifacts shall be covered below. In addition, in Fig. \ref{fig4}, we present relevant meteorological data collected from each of the substation locations during the demonstration period.

\subsection{Link 1: OC - A}
QKD system 1 (QKD1) uses the BBM92 \cite{Bennett1992} scheme, where a polarization entangled photon source at Alice is employed to generate photons in the maximally entangled singlet state $\ket{\Psi^{-}} = \frac{1}{\sqrt{2}}(\ket{H_{1},V_{2}}-\ket{V_{1},H_{2}})$. One of the photons in the entangled pair is shared with Alice, while the other photon is distributed over an optical fiber to Bob. A lithium niobate modulator at Bob ensures that any polarization fluctuations resulting from the optical fiber are corrected prior to measurement. A beamsplitter inside of both Alice and Bob randomly selects which measurement basis (either rectilinear or diagonal) to measure the received photon. When the same basis is chosen by Alice and Bob, a correlated measurement will result that is known only to Alice and Bob. These correlated measurements are used with the CASCADE \cite{Pedersen2013} error correction algorithm (with BBBSS'91 search) to produce a correlated bit-stream. A SHA-256 hash function is used for privacy amplification of the final 256-bit key shared by both Alice and Bob. The classical communications channel is authenticated using a pre-shared key.

In the course of this field test, QKD1 supplied approximately one secret key of 256-bit length every 2 seconds to the trusted node network. QKD2 and QKD3 systems generate relatively larger rates of secret key material via their respective BB84 \& SARG04 protocols. Accordingly, the NM node polls QKD1-A once per second for a new key, resulting in QKD1-B issuing the same key to TN1. The QKD1 system provided keys through a transmit-only serial communication channel. The keys were in a standardized packet format (SSP-21 / QIX) that does not report the QBER of the QKD protocol, only whether the key is "secure" or "compromised" based on an internal QBER measurement exceeding 13 \%.

\begin{figure}[!t]
\centering
\includegraphics[width=3.3in]{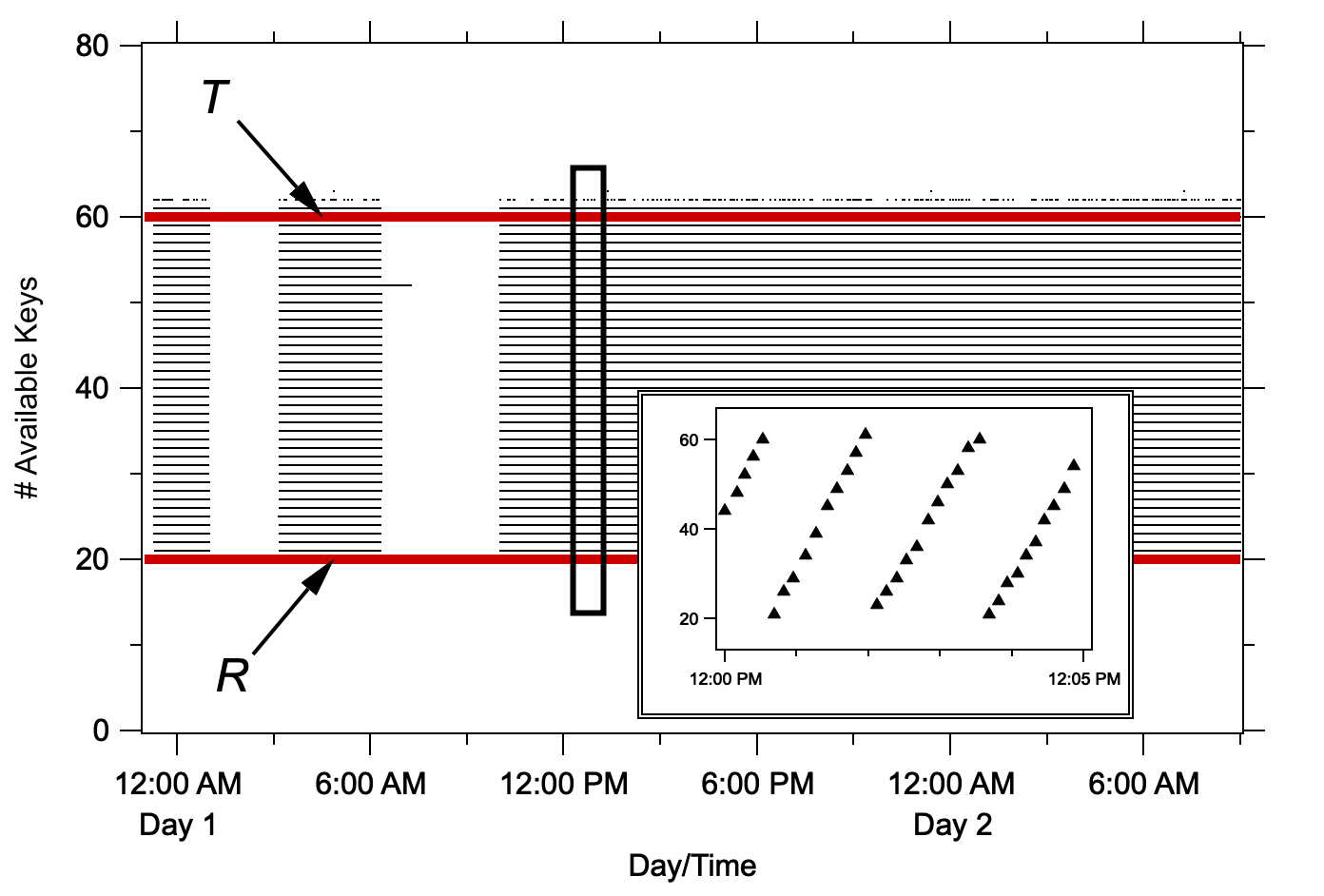}
\caption{$QK_{1}$ - the number of available quantum keys available on Link OC - A as a function of time. Also shown are the $T$ and $R$ key thresholds of 60 and 20 respectively. See text for details. The inset shows the number of available keys in a 5-minute period. The linear increase in key to $T$, followed by a quick drop to the $R$ thresholds respectively can be clearly seen.}
\label{fig5}
\end{figure}

Fig. \ref{fig5} shows the number of available quantum keys on the OC - A link ($QK_{1}$) reported by the NM and TN1 nodes. $QK_{1}$ exists within the range set by $R$ and $T$, the reserve and threshold key numbers respectively. Keys are added until $QK_{1} = T$ upon which key transfer is initiated and quantity $(T-R)$ $QK_{1}$ keys are used to transfer quantity $(T-R)$ network keys.

\subsection{Link 2: A - B}
QKD system 2 (QKD2) implements the polarization-encoded BB84 protocol \cite{Bennett2014, LANL1, LANL2, Hughes2013}. The transmitter comprises a single weak-coherent pulse (WCP) source at 1550 nm coupled into a waveguide polarization modulator and variable optical attenuator. Using a single WCP source eliminates possible side channel attacks via, for example, mismatched wavelengths or temporal pulse differences arising from multiplexed sources. The mean photon number was set to 0.1 throughout these tests; the transmitter operates at 10MHz repetition rate. A time-tracking feedback loop adjusts the detector gate delay to any position in each 100 ns time bin; it is able to accommodate clock drift and changes in the channel delay. Loss in this channel (3.1 dB) is not high enough to require a decoy-state protocol, though the hardware supports this functionality over channels with higher loss. The receiver terminal includes a polarization-tracking subsystem which corrects for polarization changes introduced by the channel (e.g. aerial fibers in the wind, temperature changes, etc.). Forward error correction is performed via low-density parity check codes at the transmitter and receiver nodes. The classical communications channel is authenticated using a pre-shared key.

One iteration of the BB84 protocol sequence constitutes a `cycle', with the end result being the creation of correlated keys. A new key file is created on both QKD2-A and -B machines upon system startup. Key material is added following every BB84 cycle and the appended file is uploaded to the TN nodes. The TN system monitors the key file size and modification date, and when triggered following a BB84 cycle, the file is reloaded to gather new key material as described earlier.

\begin{figure}[!t]
\centering
\includegraphics[width=3.3in]{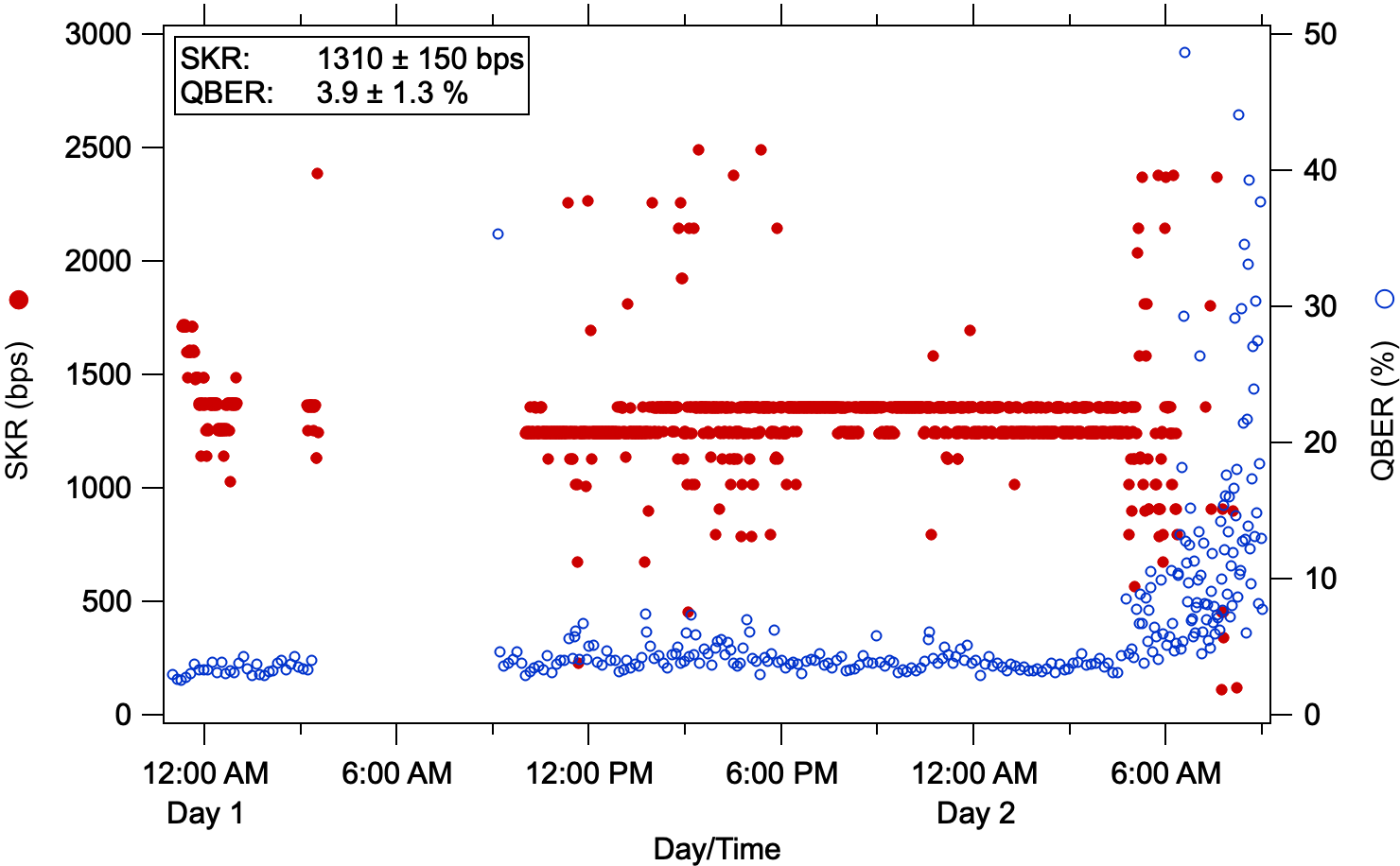}
\caption{QKD2 secret key generation rate (filled red circles - left axis) and QBER (open blue circles - right axis) on the A - B ($n$ = 2) link.}
\label{fig6}
\end{figure}

Fig. \ref{fig6} shows the SKR and QBER for Link A-B which we measure to be 1310 $\pm$ 150 bps and 3.9 $\pm$ 1.3 \% respectively. Of note is the sudden increase in QBER (and subsequent drop in SKR) beginning around 0500 on Day 2. This coincides with a period of increased wind speed, resulting in greater movement of the aerial fiber, and the polarization tracking system losing lock.

\subsection{Link 3: B - C}
\begin{figure}[!t]
\centering
\includegraphics[width=3.3in]{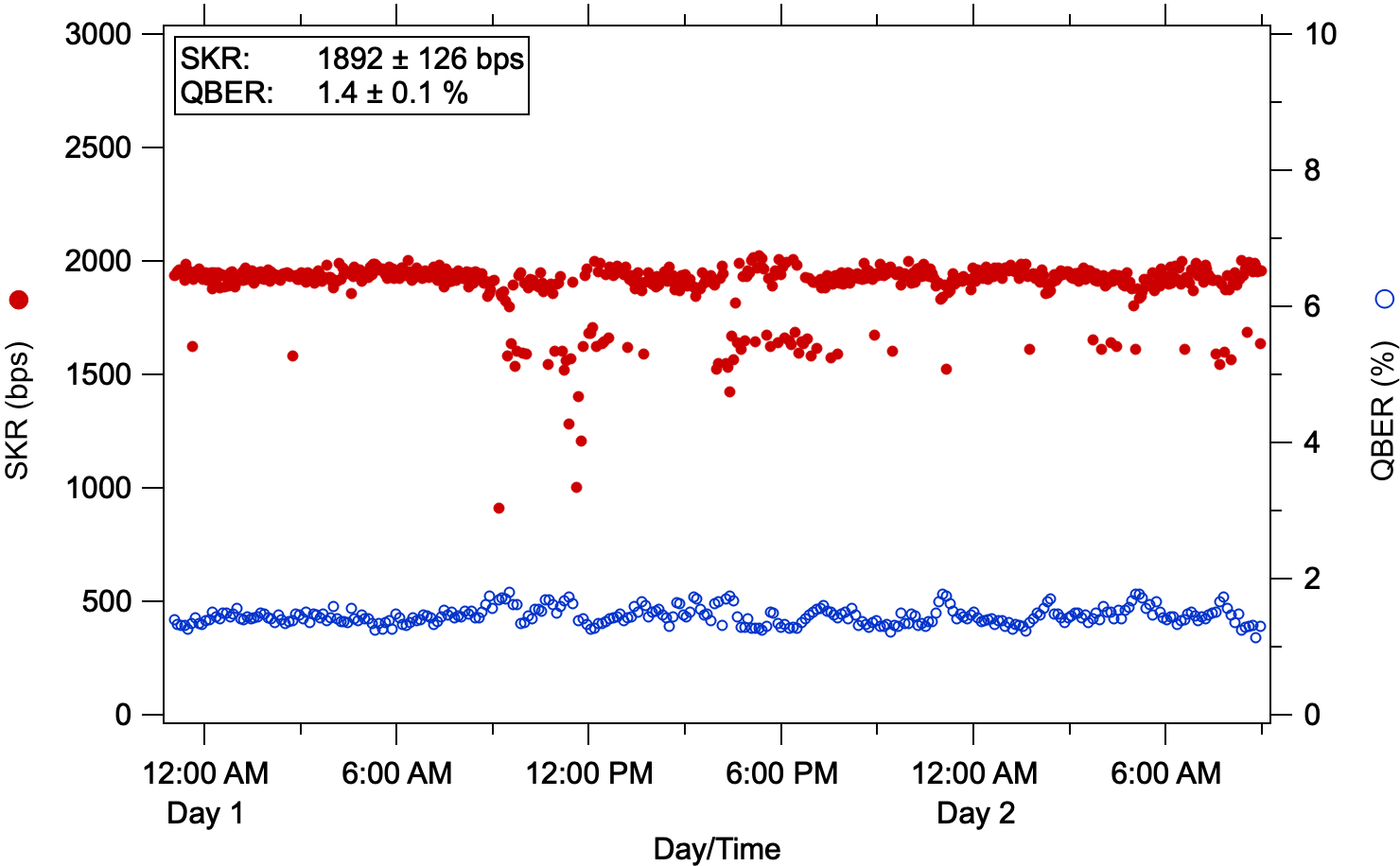}
\caption{QKD3 secret key generation rate (filled red circles - left axis) and QBER (open blue circles - right axis) on the B - C ($n$ = 3) link.}
\label{fig7}
\end{figure}

QKD system 3 (QKD3) uses a phase-encoded SARG04 protocol \cite{Scarani2004}, a variant of BB84. It proceeds by a raw key exchange phase, over the quantum channel, followed by sifting, error correction, and privacy amplification over the classical channel. The classical communications channel is authenticated using a pre-shared key. Correlated `secret' keys are saved to a file on both QKD3-A and -B upon the successful completion of each SARG04 QKD cycle. The TN system again monitors the secret key file size and modification date indicating new key material. When detected, the new key material is loaded into the TN system. Fig. \ref{fig7} shows the SKR and QBER of the QKD3 link, which we determine as 1892 $\pm$ 126 bps, and 1.4 $\pm$ 0.1 \% respectively, averaged over the demonstration period.

\subsection{Network Keys}
The growth of network keys during the field demonstration can be observed in Fig. \ref{fig8} below. Recall that the trusted node network will use quantum keys from each link upon reaching the threshold in order to generate and distribute network keys across the entire network. As a result, there is some degree of network communication overhead, resulting in the effective network key generation rate being slower than the `slowest' QKD system on the network. We demonstrate a network secret key generation rate of 115 bps, equivalent to 0.45 256-bit keys/s.

\begin{figure}[!t]
\centering
\includegraphics[width=3.3in]{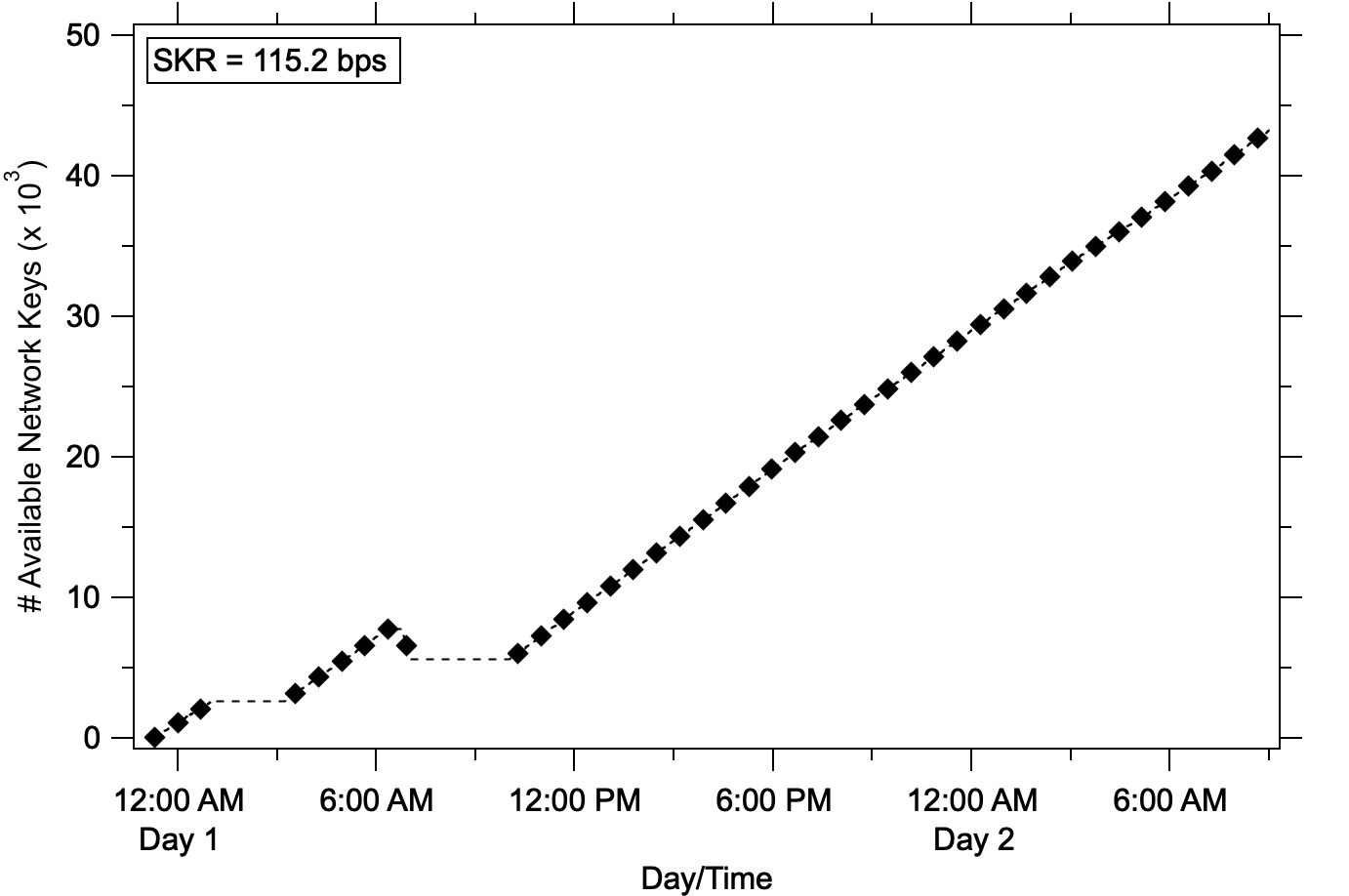}
\caption{The number of network keys across the network available for use over time.}
\label{fig8}
\end{figure}

\section{Conclusion}
\label{sect:conclusion}
To conclude, over a period of 28 hours, we performed a field demonstration of a quantum trusted node network, comprised of three different QKD systems, on the optical fiber deployed by EPB in their communications center and substations. This is the world's first such demonstration relaying keys between energy delivery infrastructure. We successfully demonstrated the interoperability of these diverse QKD systems' keys and demonstrated the generation and distribution of network keys across the network for use in critical infrastructure focused applications.

\section*{Acknowledgment}
We acknowledge the support of Steve Morrison, Tyler Morgan, and Ken Jones at the Electric Power Board (EPB), Chattanooga, TN. This work was performed at Oak Ridge National Laboratory (ORNL). ORNL is managed by UT-Battelle, LLC, under Contract No. DE-AC05-00OR22725 for the DOE. We acknowledge support from the DOE Office of Cybersecurity Energy Security and Emergency Response (CESER) through the Cybersecurity for Energy Delivery Systems (CEDS) program.

\printbibliography

\end{document}